# The Market Design for Formulary Position

by
Lawrence W. Abrams, Ph.D
March 13, 2023
Revised: September 24, 2023

**Sections:**

    **Introduction**

    **The Market for Formulary Position**

    **The Relevance of Google's Ad Position Auction**

    **Mature vs Immature Therapeutic Classes**

    **A Linear Assignment Model for Formulary Position**

    **An Illustration Using Data from Humira vs Biosimilars**

    **A Price-Cost Test for Exclusionary Lump Sum Rebates**

**Summary**

Prescription drug rebates are payments for **position** in formulary markets. The focus of this paper is the bases required to assign positions so as to minimize total expected benefit costs. We develop a linear assignment model of an "immature" therapeutic class market where there is uncertainty over a new entrant's market share.

We will show that formulary position assignments should be made on the combined bases of prices after net unit rebate bids (hereafter net unit prices) and estimates of expected demand.

The addition of lump sum rebates in the formulary position market is often judged to be exclusionary to new entrants to a therapeutic class. We use our model to develop a quantifiable test based on antitrust law standard of price - cost over a contestable market.

We view our linear assignment model as one stage in an iterative combinatorial auction featuring a menu of formulary positions with differing attributes. These attributes consist of numbers of competitors and tier copayments that have both negative and positive complementaries. What we model in this paper is an in-between round of an iterative combinatorial auction known as ["the winner determination problem"](#) where rebate bids are parameters of a prior auction round.

Future papers will model the auction round itself where rebate bids are variables in an attempt to sort out the causes of 2nd degree price discrimination rebates for position "bundles" with varying numbers of competitors.

**Introduction**

A market is an institution where buyers and sellers exchange goods and services. It operates on agreed upon rules or a design. This paper focuses on the market design for positions on a list of prescription drugs covered by health insurance plans.

We view that overall design as an iterative combinatorial auction featuring a menu of formulary positions with differing attributes. These attributes consist of numbers of competitors and tier copayments with both negative and positive complementaries. What we model in this paper is an in-between round known as ["the winner determination problem"](#) where rebate bids are parameters of a prior auction round.

In our case, the winner's determination problem involves solving a linear assignment equation with bidder market share as the unknown and rebate bids as parameters determined in the prior auction round.

**The Market for Formulary Position**

A list of covered drugs is known as a formulary. The buyers of positions are prescription drug (Rx) companies (Pharma) and the sellers are the Big 3 pharmacy benefit managers (PBMs).

This conceptualization is distinctly different from the common conceptualization of this market as a market for *quantities* of Rx drugs with Pharma as sellers and PBMs as buyers. To be sure, PBMs do contract with Pharma to buy drugs for resale to healthcare plan sponsors. But, those purchase contracts are secondary to contracts for favored position on formularies.

This conceptualization has relevance for the applicability of the Robinson-Patman Act claiming 3rd degree price discrimination because PBMs require higher rebates for a shared formulary position from new entrants to a therapeutic class. This is because the courts have determined that the Robinson-Patman applies to the sale of goods not services. Our conceptualization has PBMs as sellers of a service -- access to markets with reduced competition.

As we will discuss in more detail below, the structure of this market is that of a bilateral oligopoly with concomitant complex negotiating strategies. Available for capture from Pharma by PBMs are excess profits, or what economists call surplus or rent.

This surplus amounts to [$100s of Billions](#) accruing initially to patent-protected drugs. The reason why this surplus is up for capture is the presence of competition for formulary positions by therapeutic equivalents, which includes therapeutically equivalent patented brand drugs, biosimilars, and generics.

Formulary markets also exist for drugs administered in physicians' offices or in hospitals. These drugs are covered by medical benefit plans managed by insurance companies and not PBMs. Both PBMs and insurance companies are agents hired for their expertise and size in negotiating lower drug benefit costs.

Because of the oligopolistic structure of this market, both buyers and sellers have discretion over the choice of bases of exchange. There are multiple bases in this market. They vary from therapeutic class to therapeutic class and change over time. They also depend on the business model of the benefits manager.

The money exchanged for favored positions are called drug rebates. Drug rebates have been given a safe harbor by Federal and State healthcare anti-kick back laws. In almost all cases, drug rebates flow off-invoice. This is in contrast to on-invoice discounts passed through the supply chain as incentives for volume purchase efficiencies and prompt payments. Unfortunately, it is rare to find articles on the drug supply chain that make a clear distinction between drug rebates and drug discounts.

Rebate negotiations start with a flat % off list prices as measured by a publicly available standard called wholesale acquisition costs (WAC). This basis is verified in the [extensive US Senate investigation](#) into insulin rebates negotiations.

> Rebates are payments made by drug manufacturers to PBMs after the point of sale,[183] and are calculated as a percentage of WAC. Drug manufacturers negotiate rebates with PBMs and health insurers to secure preferred formulary placement for their products.[184] These negotiations can be of such great financial importance to pharmaceutical companies that senior executives up to and including the chief executive officer are often personally involved in the process.[185]

There are a multitude of other bases that could be added to negotiations for formulary positions such
1. Lump sums rebates
2. Rebates tied to a bundle of drugs
3. Rebates tied to copayment tiers on the formulary
4. Rebates tied to usage limitations like prior authorization
5. Rebates tied to drug indications and not the drug itself
6. Rebates tied to outcomes (e.g. cancer drugs rebates paid yearly for each additional year of progression free survival)

The extent of the possibilities listed above is much greater than that found typically in oligopolistic markets for physical goods. This is a reflection of a market for something ephemeral -- **positions** providing access to markets with reduced competition.

The structure of the market for formulary position is that of a bilateral oligopoly. The sell side is dominated by three large PBMs [controlling 80% of the total Rx claims](#) processed in the United States in 2021. The so-called Big 3 are all now vertically integrated — Express Scripts (Cigna), Caremark (CVS Aetna), and OptumRx (United Healthcare).

Records show that hospitals first used the word [formulary in the 1800s](#) to describe lists of drugs in their dispensaries accompanied by standard operating procedures for usage. In the 1980s, hospitals began to use formularies for cost management by steering usage to generics coming into the market as perfect substitutes for more costly off-patent brand drugs.

PBMs started out as computer networking specialists who automated Rx claims processing by connecting pharmacy point of sales terminals to back-office health insurance mainframes. Between 1980-1990, PBMs' primary source of revenue was claims processing fees. Following hospitals, PBMs added a look-up table to the point of sale software in the 1990s. This allowed for automatic switching of prescriptions for off-patent brands to their much lower cost generics.

Starting in 2000, the most popular therapeutic classes of drugs -- proton pump Inhibitors, COX-2 inhibitors, 2nd generation antihistamines, and statins -- started to see the entry of therapeutic equivalents with a mere entamer difference in molecular structure. The opportunity for capturing some of the excess profits generated by patent-protected drugs was an order of magnitude greater than surplus capture from switching off-patent brands to generics. Give PBMs credit for realizing that formularies could be used to create competition among drugs that otherwise were patent protected monopolists.

Our 2005 paper [PBMs as Bargaining Agents](#) was the first to apply economics to the question of why rebates were paid and where. We conceptualized therapeutic classes within formularies as a group of markets. Since substitutability is a key structural feature defining competition within any market, we saw therapeutic classes as markets with varying degrees of competition for a favored position. We conceptualized rebates as tariffs paid to PBM gatekeepers for entry to markets with reduced competition. We broke down therapeutic class markets into three types:

1. competitive — featuring at 4+ drugs that have lost patent protection and have lower cost generics that are therapeutic equivalents;
2. monopolistic – featuring a single first-to-market "innovative" patented drug;
3. oligopolistic — featuring a small number of patented drugs and 1-2 generic or biosimilar drugs that are therapeutic equivalents.

Our conclusion was that rebate negotiations only took place in oligopolistic therapeutic classes. In addition, the locus of the greatest rebate dollars changed over time as the competitiveness within therapeutic classes changed. From past papers, here are a couple of snapshots of specific therapeutic class bilateral oligopolies showing the variability of position assignments by PBM by year.

From the paper: [Insulin Drug Price Inflations: Racketeering or Perverse Competition?](#)

### 2018 National Formulary for Long-Lasting Insulin Therapeutic Class

| Pharma Co. | Relation To Incumbent | FDA Approved | Drug | CVS Health Included | CVS Health Excluded | Express Scripts Included | Express Scripts Excluded | Prime Therapeutics Included | Prime Therapeutics Excluded | OptumRx Included | OptumRx Excluded |
|---|---|---|---|---|---|---|---|---|---|---|---|
| Sanofi | Incumbent | 4-2000 | Lantus® | | x | x | | x | | x | |
| **Eli Lilly** | **Follow-on Biologic** | **12-2015** | **Basaglar®** | **x** | | | **x** | | **x** | | **x** |
| Novo Nordisk | Therapeutic Equiv | 6-2016 | Levemir® | x | | x | | x | | | x |
| Novo Nordisk | Therapeutic Equiv | 12-2016 | Tresiba® | x | | x | | x | | | x |
| Sanofi | Therapeutic Equiv | 2-2015 | Toujeo® | | x | x | | x | | x | |

Both below from the paper: Was CVS' Exclusion of Mavyret a Violation of Antitrust Law?

### 2017 Formulary Choice for Hepatitis C Virus Class - Express Scripts vs CVS

| List Price Regimen | Manufacturer | Drug | CVS - 2017 | ESRX - 2017 | FDA Approved | 1 pill a day weeks | Genotype |
|---|---|---|---|---|---|---|---|
| $ 84,840 | Gilead Sciences | Sovaldi | | Excluded | 12-6-13 | 12 | 1,2,3,4 |
| $ 95,445 | Gilead Sciences | Harvoni | Preferred | | 7-10-14 | 12 | 1,4,5,6 |
| $ 75,508 | Gilead Sciences | Epclusa | Preferred | | 6-28-16 | 12 | 2,3 |
| $ 21,038 | AbbVie Inc | Viekira Pak | Excluded | Preferred | 12-19-14 | 12 | 1 |
| $ 38,710 | AbbVie Inc | Technivie | Excluded | Preferred | 7-24-15 | 12 | 4 |
| $ 67,024 | Janssen T. | Olysio | Excluded | Excluded | 11-22-13 | 12 | 1,4 |
| $ 63,630 | B-M-S | Daklinza | Excluded | Excluded | 7-24-15 | 12 | 1,3 |
| $ 55,148 | Merck | Zepatier | Excluded | Excluded | 1-28-16 | 12 | 1,4 |
| $ 74,760 | Gilied Sciences | Vosevi | not FDA | not FDA | 7-18-17 | 12 | 1-6 |
| $ 26,400 | **AbbVie Inc** | **Mavyret** | not FDA | not FDA | 8-3-17 | 8 | 1-6 |

### Express Scripts Formulary Choice for Hepatitis C Virus Class - 2017 vs 2018

| List Price Regimen | Manufacturer | Drug | ESRX - 2017 | ESRX - 2018 | FDA Approved | 1 pill a day weeks | Genotype |
|---|---|---|---|---|---|---|---|
| $ 84,840 | Gilead Sciences | Sovaldi | Excluded | Excluded | 12-6-13 | 12 | 1,2,3,4 |
| $ 95,445 | Gilead Sciences | Harvoni | | Preferred | 7-10-14 | 12 | 1,4,5,6 |
| $ 75,508 | Gilead Sciences | Epclusa | | Preferred | 6-28-16 | 12 | 2,3 |
| $ 21,038 | AbbVie Inc | Viekira Pak | Preferred | Preferred | 12-19-14 | 12 | 1 |
| $ 38,710 | AbbVie Inc | Technivie | Preferred | Preferred | 7-24-15 | 12 | 4 |
| $ 67,024 | Janssen T. | Olysio | Excluded | Excluded | 11-22-13 | 12 | 1,4 |
| $ 63,630 | B-M-S | Daklinza | Excluded | Excluded | 7-24-15 | 12 | 1,3 |
| $ 55,148 | Merck | Zepatier | Excluded | Excluded | 1-28-16 | 12 | 1,4 |
| $ 74,760 | Gilied Sciences | Vosevi | not FDA | Preferred | 7-18-17 | 12 | 1-6 |
| $ 26,400 | **AbbVie Inc** | **Mavyret** | not FDA | Preferred | 8-3-17 | 8 | 1-6 |

Our paper The Three Phases of the PBM Business Model summarized our financial disaggregation of PBM 10-Qs and 10-Ks showing how the distribution of PBMs' gross profits changed in response to changes in the nature of competition within therapeutic classes.

We summarize below changes we saw in market design in response to changes in the nature of competition and the need by PBMs for new sources of retained rebates:

1. Our 2017 paper blamed PBMs for changing the market design to a two step high list - high rebate bidding process resulting in a gross-to-net drug price bubble. We traced that change to a need to replace declining gross profits from their mail order operations and a shift in available surplus from small molecule brands to large molecule biologics.

2. We traced the growth after 2012 in the number of drugs outright excluded from formularies to the need by PBMs to capture more rebates to offset losses from small molecule classes losing patent protection like statins, 2nd generation antihistamines, and proton pump inhibitors.

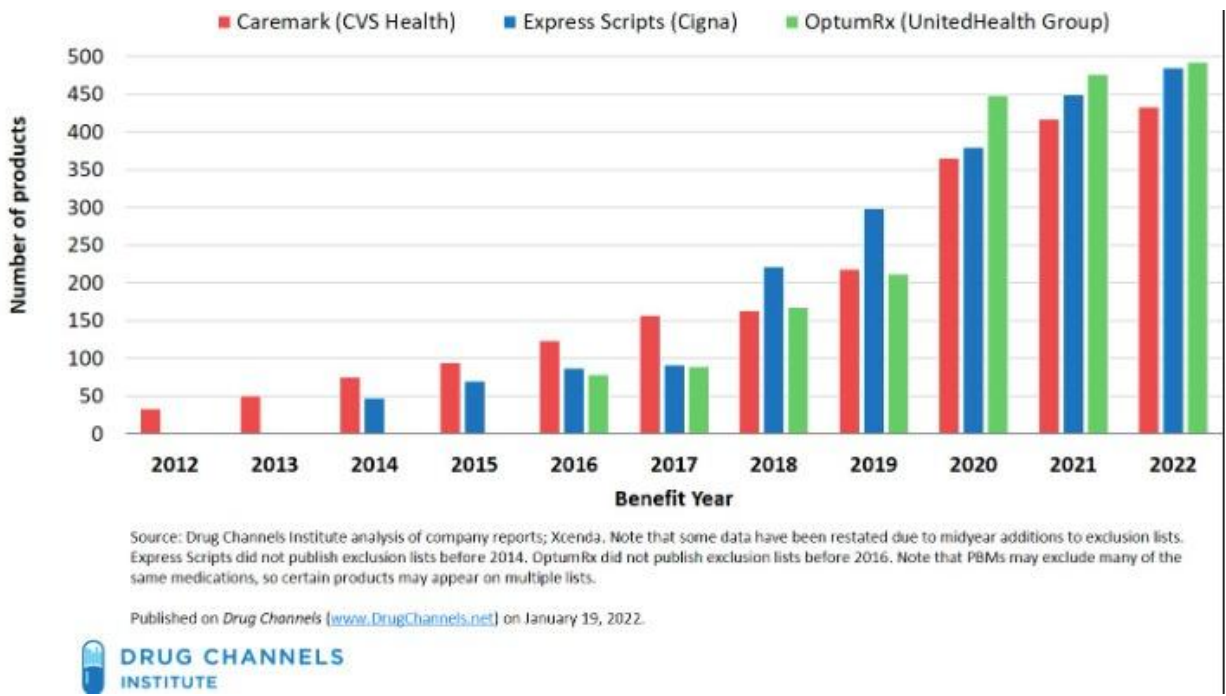

3. Because biologic drugs in therapeutic classes like autoimmune diseases and oncology are approved by the FDA for select indications, PBMs have recently subdivided these therapeutic classes into subclasses like psoriasis, spondylitis, and psoriatic arthritis. By creating indications-based formularies, PBMs created new competition for positions and new opportunities for surplus capture. A future paper will show how a combinatorial auction is ideally suited to capture the complementaries in assigning positions in an indication-based formulary and produce "shadow prices" for individual indications.

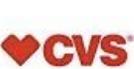

Figure 4: Cosentyx vs. Taltz on Indication-based Formulary, 2020-2021

|  |  | AS | nr-axSpA | PsO | PsA |
|---|---|---|---|---|---|
| CVS | 2020 | Cosentyx (secukinumab) | No indication-based formulary | taltz (ixekizumab) injection 80 mg/mL | Cosentyx (secukinumab) |
| CVS | 2021 | taltz (ixekizumab) injection 80 mg/mL | No indication-based formulary | taltz (ixekizumab) injection 80 mg/mL | Cosentyx (secukinumab) |
| EXPRESS SCRIPTS | 2020 | Cosentyx (secukinumab) | | | |
| EXPRESS SCRIPTS | 2021 | taltz (ixekizumab) injection 80 mg/mL | | | |

Source: ESI/CVS 2021 Formularies

4. We believe that an important motivation for PBMs to integrate vertically with insurance companies five years ago was an accelerating shift in available surplus capture from large molecule biologics covered by medical benefit plans. Below is an unpublished graph compiled by Alex Telford @atelfo on Twitter showing the relative trend in the number of blockbuster drugs in each class.

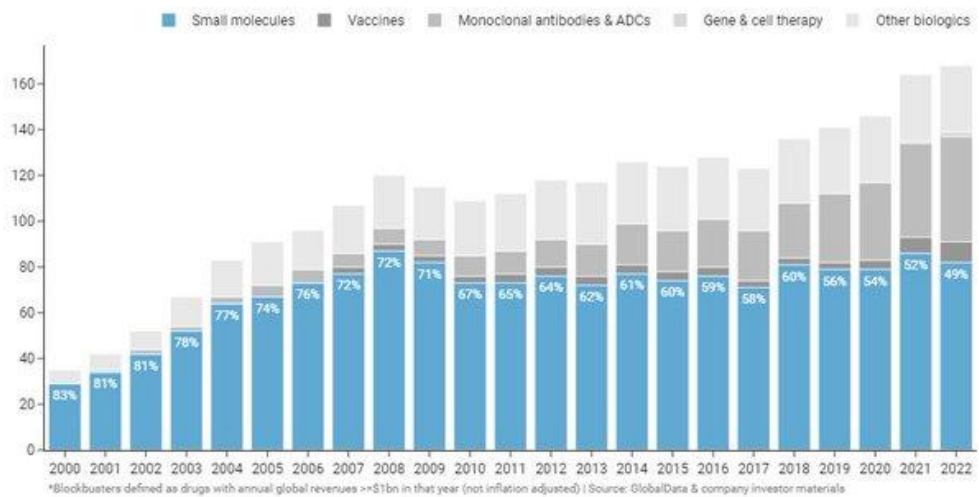

Because rebate contracts contain sensitive corporate information that could damage a company's competitiveness, there are plenty of government regulations and corporate contract clauses preventing disclosure. Most data on rebates comes from government reports. For example, The United States Office of Inspector General periodically audits PBMs as Medicare Part D plan sponsors and discloses generalized information about rebate averages and bases.

By far, the most detailed disclosure of rebate negotiations and the bases used comes from a 2019 Grassley-Wyden Senate Staff Report investigating the insulin drug bilateral oligopoly involving Sanofi, Novo Nordisk and Eli Lilly as buyers of position and the Big 3 PBMs as sellers. This report was based on over 100,000 pages of internal documents.

**The Relevance of Google's Ad Position Auction**

From the outset, it was clear to Google that advertisers were willing to pay more for banner ads on the first search page and more for side-bar ads at top of any given search page. Given the dynamic nature of on-advertising, fixed contracts based on volume discounts were not feasible. The basis for bids had to be a simple $ per click that changed frequently.

The problem with an ad auction where assignment is based solely on $ per click was that Google's revenue is the PRODUCT of $ per click bids times click-through rates. Assigning ad positions only on $ per click could result in too many top positions going to high unit bidders with ads having little appeal and low click-through rates.

It was Google's chief economist Hal Varian who introduced Google to the [economics of position auctions](#) where the winner's assignment problem entailed an equation that maximized total expected revenue combining $ per click and an estimate of click-through rates. It was Google's co-founder Larry Page who developed a [complex estimate](#) of expected click-through rates called AdRank. Here is a link to a [video of Google's chief economist Hal Varian](#) presenting an example of how Google's position auction works. In the example, the top ad position went to the bidder with a relatively low unit bid but a very high AdRank.

The need by Google to go beyond simple $ per unit bids is similar to the need for PBMs to go beyond assigning formulary positions solely on the basis of net unit prices. Because the objective of a PBM is to minimize total benefit costs, this requires them to assign formulary positions on the combined bases of net unit prices and estimates of expected demand.

**Mature vs Immature Therapeutic Classes**

We find it useful to divide the formulary position market into two types of therapeutic classes -- "mature" and "immature" therapeutic classes.

In mature therapeutic classes, a PBM can draw on a long history of therapeutic equivalents switching in and out of favored positions. Physicians are familiar with patients' experiences switching among therapeutic equivalents. Competitors have proven their capability of filling demand if given an exclusive or a dual shared position.

Mature therapeutic classes still involve positions with negative and positive complementaries due to varying numbers of competitors in a position and tier copays. An iterative combinatorial auction still is the best market design to capture these complementaries. There still will be 2nd degree price discrimination due to position attribute differences.

What we think will be different is the absence of 3rd degree price discrimination on the basis of expected demand of any drug. Concern about expected demand is absent because any single drug in a mature therapeutic class is capable of filling all demand. In the language of Google ad auctions, the click-through rate is the same for any drug if given a particular position.

In the next section, we develop a linear assignment equation for optimal assignment of formulary positions in immature therapeutic classes. This case is interesting for two reasons. Because the expected market share of the new entrant is uncertain, the PBM's objective becomes more complex than simply assigning positions based solely on net prices.

Second, it is the uncertainty over an entrant's market share that is the source of an incumbent's bargaining power. Nominally it is a PBM who adds lump sum rebates as bid requirement. But, the initiator of such a move is likely the incumbent. A PBM cannot afford to lose the incumbent due to the inability of the entrant to meet all demand.

While immature therapeutic classes might be relatively small in number, the potential for new surplus capture in each case is many times that in mature therapeutic classes. We specifically base our model on the 2023 case of AbbVie's $21 Billion blockbuster biologic Humira facing its first biosimilar competitor from Amgen's biosimilar Amjevita. Throughout the rest of this paper, we will be using published estimates from that case to

present bid menus and graphs illustrating the trade-offs a PBM faces in assigning formulary positions.

**A Linear Assignment Model for Formulary Position**

As stated earlier, the market design for formulary position is similar to Google's ad position auction market design.  In a 2011 NBER working paper, Levin has noted that Google's market design is a special case of the classical assignment model of Shapley and Shubik.  As noted by Shapley and Shubik, their assignment problem can be cast as a linear programming model.  We conceptualize a PBM's role in the formulary position market using the language of both linear programming model and the assignment problem.

We view that market design for formulary position as an iterative combinatorial auction featuring a menu of bundled packages with different attributes -- positions with varying numbers of competitors and tier copayments.  What we model here is an in-between round known as "the winner determination problem"  with rebate bids as parameters of a prior auction round.

In our case, the winner's determination problem is solved by a linear assignment equation with bidder market share as the unknown and rebate bids as parameters determined in the prior auction round.

Future papers will model the auction round in an attempt to sort out the causes of multiple 2nd and 3rd degree price discrimination presented in menus below. Evidence of these price differences can be found in the 2019 extensive U.S. Senate investigation into insulin pricing.

The model features a single PBM tasked with assigning two possible positions: a shared position and an exclusive position.  There are only two competitors: an entrenched biologic incumbent just coming off patent protection versus the first biosimilar entrant.

To keep our model simple, we assume that if the entrant isn't assigned the shared position, it is relegated to an inferior formulary position like fail first or prior authorization such that its resultant market share is immaterial to the total market managed by the PBM. In the language of assignment problems, we keep the model a balanced assignment problem by pairing the exclusive assignment with a "dummy" assignment with zero benefit costs.

The objective of a PBM is to make a shared formulary position assignment only if the expected benefit costs are less than or equal to the certain benefit costs of an exclusive assignment to the incumbent.

In algebraic terms, the dependent variable is total expected benefit costs. The independent variable is market share. The parameters are unit bids % off lists for positions from a prior iteration of a combinatorial auction of a menu of formulary positions of varying numbers of competitors with negative complementaries.

We believe these negative complementaries stem from varying costs of residual non-price competition remaining in a formulary position after excluding other competitors. Evidence supporting this hypothesis is the focus of a future paper.

The exogenous variables are list prices and total Rx market in units. For simplicity we assume list prices are the same for both competitors even though in most cases the entrant lists its price around 5% less.

| | | Rebate bids as % off List | |
|---|---|---|---|
| List price | | Exclusive | Shared |
| Z | Incumbent | b1 | b2 |
| Z | Entrant | | b3 |

Exogenous variables: T = Total market in units, Z = List price

Dependent Variable is total expected benefit cost (TBC)
Independent variable = x market share of entrant

| | | Y Intercept b | Slope a |
|---|---|---|---|
| Linear Equation: | $Y = ax + b$ | TBC when $x=0$ | |
| TEBC - Exclusive | $= (T*Z)*(1- b1)$ | $TZ*(1-b1)$ | 0 |
| TEBC - Shared | $= (T*Z)*(1- b2)*(1-x) + (T*Z)(1-b3)*x$ | $TZ*(1-b2)$ | $-T*Z*(b3-b2)$ |
| | $=(T*Z)*(1-b2) - [(T*Z)*(b3-b2)]*x$ | | |

Slope - a % increase in market share of entrant reduces TBC $ by $-T*Z*(b3-b2)*.01$

Market Share of Entrant Required to Equalize TEBC

$$x = \frac{(b1 - b2)}{(b3 - b2)}$$

where (b1 - b2) is the incumbent's position bid down differential > 0
where (b3 - b2) is shared position bid differential - entrant vs incumbent > 0

If both agents had the same certain capability of filling 100% of Rx demand, the assignment would be easy. Assign exclusive position to the agent with the lowest net unit price.

The formulary assignment problem becomes complicated when it is uncertain whether the entrant can meet any expectations for market share. This uncertainty is a source of an incumbent's bargaining power because it knows the PBM cannot give an entrant an exclusive position no matter how low the entrant's net price bid is. In other words, the PBM has significantly different elasticity of demand for the two buyers and exploits this by demanding higher rebate %'s from the entrant relative to the incumbent for the same shared formulary position.

Even if the entrant offers its drug for free, a PBM still needs the incumbent to fill most of the demand. As a result, the incumbent increases its own bid spread between an exclusive assignment and a shared assignment. And depending on that spread, it is possible that an exclusive assignment has a lower expected benefit cost than a shared assignment even with the entrant's net price at zero.

Here is a simple numerical example of an assignment problem where there is uncertainty over one competitor's capabilities. Start with an incumbent bidding $50 per hour for an exclusive 100 hour work assignment. Total bid is $5,000. For a shared assignment, the entrant bids $30 per hour but cannot guarantee anything about degree of completion. The incumbent knowing it might have to jump in and "mop up" if the entrant falls short, raises its shared bid to $80 per hour. The problem facing the decision-maker desiring to minimize the cost of the task: exclusive assignment or shared assignment?

The decision-maker can make an informed assignment by calculating break-even hours between an exclusive assignment and a shared assignment via this algebraic equation:

> Solve for x where x hours work by entrant sufficient to break-even
> between an exclusive and shared assignment:
> 5,000 = (30*x) + (80*(100-x) )
> 80*x - 30*x = 8,000 - 5,000
>   x = 60 hours

If the decision-maker believes the entrant is likely to work 60 hours or more, it should make a shared assignment, otherwise assign the task exclusively to the incumbent.

**An Illustration Using Data from Humira vs. Biosimilars**

To add some relevance and insight into our model, we apply it to the 2023 case of AbbVie's blockbuster drug Humira finally losing patent protection after 20 years. Indeed, Humira could be the blockbuster of all-time blockbuster drugs with $20.7 Billion in 2021 US sales alone. The sell side in our model is one of the Big 3 PBMs Because this story involves unprecedented potential surplus capture from a drug finally losing patent protection,  there are plenty of available estimates to draw from publications like [Endpoint News](), [Drug Channels]() and the [American Enterprise Institute.]() We can even draw from [AbbVie's own forecast of a 37% decline i]()n 2023 sales representing a mix of net price and quantity declines.

We use 3.5 Million self-injectable pens as the total fixed quantity of demand.  This was derived from a reported 2021 Humira [total US sales of  $20.7 Billion]()  divided by a 2021 estimate of unit list price of $1,592 times the market share of a single PBM =  ⅓ * 80% = 27% of the total US market.

Below is a list of exogenous variables for list prices and total quantity demand. Also, included is a menu of reasonable estimates for the parameters for % off  unit rebates and net unit prices.  We make it clear that the entrant has the lower net price drug after unit rebates. This is due to 3rd degree price discrimination against the entrant because the PBM's need for their inclusion is quite elastic relative to its need for the incumbent to fill most demand.

Based solely on the data below, one might conclude falsely that a shared position is best from the standpoint of minimizing total benefit costs. However, market share is a key variable that must be considered.  Depending on market share, either an exclusive assignment or a shared assignment would minimize total expected benefit costs.

| T = 3.5 Million Rx | | Rebate as % off List | |
| --- | --- | --- | --- |
| WAC $/Unit | | Exclusive | Shared |
| $1,731 | Incumbent | 50% | 44% |
| $1,731 | Entrant | | 70% |

| T = 3.5 Million Rx | | Net Unit Prices | |
| --- | --- | --- | --- |
| WAC $/Unit | | Exclusive | Shared |
| $1,731 | Incumbent | $866 | $969 |
| $1,731 | Entrant | | $519 |

| | | |
| --- | --- | --- |
| TEBC - Exclusive ($M) | = | $3,029 |
| TEBC - Shared ($M) | = | $3,393 - $1818*x |

Slope - a 1% increase in market share of entrant reduces TBC $ by $18.18 Million

Market Share of Entrant Required to Equalize Total Benefit Costs

$$x = \frac{(b1 - b2)}{(b3 - b2)} = \frac{6\%}{26\%} = 23\%$$

We can use these estimates to illustrate an important point about rebate negotiations in an immature therapeutic class. There is so much an entrant can do to lower total benefit costs via its own high % off list bids due to its minor expected market share. Even though the entrant greatly outbids the incumbent for a shared position, a PBM knows that a shared assignment might not lower total benefit costs if the entrant does not gain a sufficient market share.

The graph below illustrates this point. It is a graph of total benefit costs as a function of the market share of the entrant assuming three different values for incumbent % off bid for a shared position. The flat blue line represents the total benefit costs of an exclusive position assignment. The graphs illustrate that a mere 10 percentage point bid down by the incumbent -- from 48% off list to 38% off list -- increases the equalizing market share of the entrant from 20% market share to 40% market share.

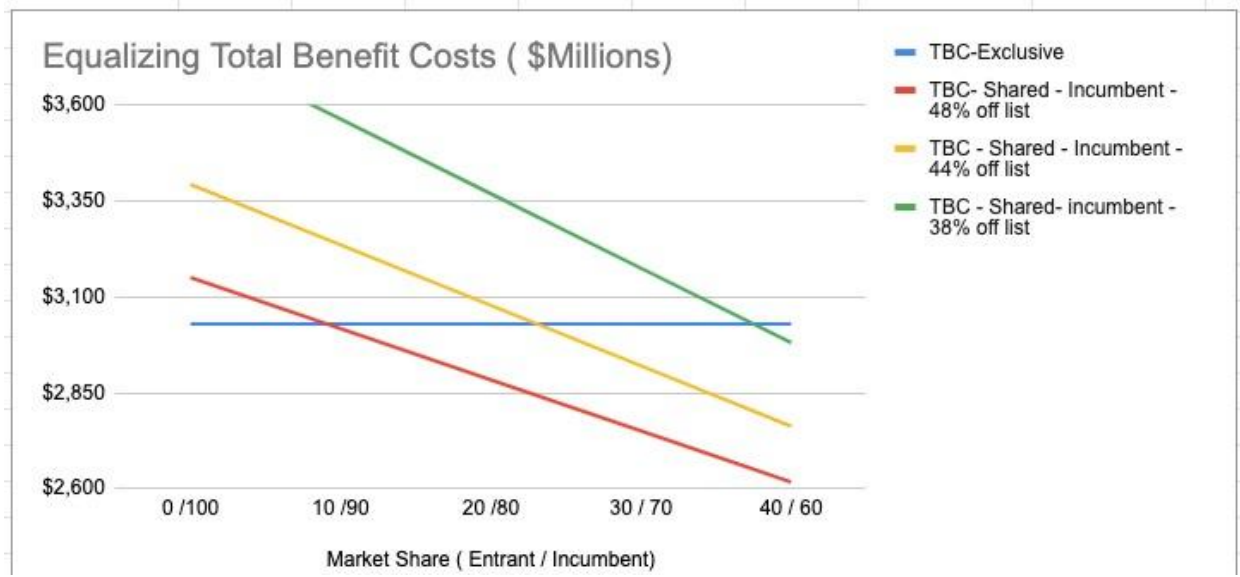

The array below is another way to illustrate this point. It is an array of market shares required by the entrant to equalize total benefit costs under various assumptions about the bid spreads. The horizontal axis is the bid down spread of the incumbent from an exclusive to a shared position. It is ($b_2 - b_1$) in the algebraic model above. The vertical axis is the bid spread of the entrant versus the incumbent for the shared position. It is ($b_3 - b_2$) in the algebraic model above. Assuming a reasonable expectation of an initial 20% market share for the entrant, the table below illustrates how difficult it is for an entrant to lower total expected benefit cost when the incumbent's biddown spread exceeds six percentage points.

Taking another look at the algebraic equation of total expected benefit costs illustrates how a bid drawdown by an incumbent can be exclusionary. In algebraic terms, an incumbent, low bid for a shared position-- $b_2$ -- increases in the Y-intercept of the equation, which translates into a higher starting place for an entrant to make its impact via market share gains. At the same time, the incumbent's shared position bid -- $b_2$ -- lowers the slope of the benefit cost equation, which means a greater marginal benefit for each 1% gain in market share by an entrant.

TEBC - Shared  $=(T*Z)*(1-b2) - [(T*Z)*(b3-b2)]*x$

Bid Spreads that Equalize Expected Benefit Costs at 20% or Less Entrant Market Share

| Entrant - Incumbent Shared % off List Bid Spreads (b3 - b2) | Incumbent Bid-Down % off List Spread -- from Exclusive to Shared (b2 - b1) | | | | | | | |
|---|---|---|---|---|---|---|---|---|
| | 3% | 4% | 5% | 6% | 7% | 8% | 9% | 10% |
| 20% | 15% | 20% | 25% | 30% | 35% | 40% | 45% | 50% |
| 25% | 12% | 16% | 20% | 24% | 28% | 32% | 36% | 40% |
| 30% | 10% | 13% | 17% | 20% | 23% | 27% | 30% | 33% |
| 35% | 9% | 11% | 14% | 17% | 20% | 23% | 26% | 29% |
| 40% | 8% | 10% | 13% | 15% | 18% | 20% | 23% | 25% |
| 45% | 7% | 9% | 11% | 13% | 16% | 18% | 20% | 22% |
| 50% | 6% | 8% | 10% | 12% | 14% | 16% | 18% | 20% |
| 55% | 5% | 7% | 9% | 11% | 13% | 15% | 16% | 18% |

**A Price-Cost Test for Exclusionary Lump Sum Rebates**

We conclude the paper with an examination of the use of lump sum rebates as an added basis for position assignment. While lump sum rebates are imposed nominally on all competitors, we will argue that it is only an incumbent that has both the motivation and negotiating power to make a PBM add them as a precondition for position assignments.

Generally accepted accounting principles would treat lump sum rebates paid by competitors as a reduction in **revenue**. That is how we present rebates in tables showing their effect on **average net prices.**

However, from an economics perspective, it is useful to view them as a **fixed cost** requirement to enter a market.  From an antitrust law perspective, they can be analyzed using the theories of raising a rival's costs.   Consistent with that view, we develop an exclusionary standard based on a price - cost test.  As we will show below with Humira case data, the imposition of lump sum rebates can drive an entrant's **average net prices** below cost of goods sold.

First, we add lump sum rebates to our algebraic model and recognize that competitors facing lump sum rebates adjust down their % off  unit rebate bids.

| Parameters | b1, b2, b3 = % off list price bids |
|---|---|
| Exogenous variables: | LS1, LS2, LS3 = lump sum rebates |
| Exogenous variables: | T = Total market in units, Z = List price |

| List price | | Rebate bids and lump sums | |
|---|---|---|---|
| | | Exclusive | Shared |
| Z | Incumbent | b1 | b2 |
| | | LS1 | LS2 |
| Z | Entrant | | b3 |
| | | | LS3 |

Below are the equations for **average net unit prices** after rebates for the incumbent and the entrant.  In the case of no lump sums, average net prices are constant across all market shares.  But, in the case of lump sum rebates, averages vary with market shares and go in opposite directions as the market share of the entrant increases.

|  | Shared Position Average Net Prices: | | | | |
|---|---|---|---|---|---|
| no lump sum Incumbent | =T*Z*(1-b2)*(1-x) / T*(1-x) | = Z*(1-b2) | Net unit price - flat | | |
| no lump sum Entrant | =T*Z*(1-b3)*x / T*x | = Z*(1-b3) | Net unit price - flat | | |
|  | Shared Position Average Net Prices: | | | | |
| lump sum Incumbent | =T*Z*(1-b2)*(1-x) - LS2 / T*(1-x) | = Z*(1-b2) - | LS2 / T*(1-x) | declines with x | |
| lump sum Entrant | =T*Z*(1-b1)*x - LS3 / T*x | = Z*(1-b1) - | LS3 / T*x | increases with x but can be < 0 | |

Most importantly, due to high fixed costs, average net prices will be negative for the entrant at low market shares. While we do not have estimates for cost of goods sold as a % off list, we would expect that gross margins become unsustainably low when average net prices exceed 80% off list prices. Obviously, gross margins are negative when unit rebates exceed 100% off list prices.

To apply a price-cost test for exclusionary lump sum rebates, one must first establish a reasonable market share performance for a new entrant if assigned a shared position. If this is not established, then an absurdly low threshold like 5% market share with corresponding negative gross margins will produce a false positive test. At a 5% market share performance threshold, negative margins are the fault of the entrant not the lump sum rebates.

Once a performance threshold is established, say 20% market share, the equation above can be used to calculate average net prices as a function of market share. A lump sum rebate would be exclusionary if the entrant's calculated average net price exceeds 80% off list price.

We can illustrate this price-cost test for exclusionary lump sum rebates using data from the Humira case. We use the same exogenous variables for total market quantity and list prices.

Because of the imposition of a lump sum rebates, both competitors adjust down their unit rebate bid as % off list.

| T = 3.5 Million Rx WAC $/Unit | | Rebate bids and lump sums | |
| --- | --- | --- | --- |
| | | Exclusive | Shared |
| $1,731 | Incumbent | 25.3% | 12.6% |
| | $ Millions | $1,500 | $1,200 |
| $1,731 | Entrant | | 30% |
| | $ Millions | | $850 |

We plug in these new rebate parameters to the equations above and calculate average net price as a function of market share. The result is that at a reasonable market share of 20%, the entrants average net price would be 100% off list -- zero -- at the 20% market share threshold.

Moreover, the new entrant would have to cross the 35% market share threshold just to achieve sustainable margins. Clearly, based on a price-cost test, an $850M lump sum rebate would be exclusionary in this case.

| Average Net Prices | | | | exclusionary | | | |
| --- | --- | --- | --- | --- | --- | --- | --- |
| Market share - Incumbent | 90% | 85% | 80% | 75% | 70% | 65% | 60% |
| Market share - Entrant | 10% | 15% | 20% | 25% | 30% | 35% | 40% |
| List Price - Entrant | $1,731 | $1,731 | $1,731 | $1,731 | $1,731 | $1,731 | $1,731 |
| Average Net Price - 70% off list / no Lump Sum | $519 | $519 | $519 | $519 | $519 | $519 | $519 |
| Average Net Price - 30% off list - $850M Lump Sum | -$1,217 | -$407 | -$3 | $240 | $402 | $518 | $605 |
| | | | | | | | |
| Net Price as % off list - no lump sum | 70% | 70% | 70% | 70% | 70% | 70% | 70% |
| Net Price as % off list - $850M lump sum | 170% | 124% | 100% | 86% | 77% | 70% | 65% |
| List Price - Incumbent | $1,731 | $1,731 | $1,731 | $1,731 | $1,731 | $1,731 | $1,731 |
| Average Net Price - 44% off list / no Lump Sum | $969 | $969 | $969 | $969 | $969 | $969 | $969 |
| Average Net Price - 12.6% off list - $1,200M Lump Sum | $1,132 | $1,110 | $1,084 | $1,056 | $1,023 | $985 | $941 |
| | | | | | | | |
| Net Price as % off list - no lump sum | 44% | 44% | 44% | 44% | 44% | 44% | 44% |
| Net Price as % off list - $850M lump sum | 35% | 36% | 37% | 39% | 41% | 43% | 46% |

We purposely chose the rebate parameters in this case to cause the break-even total benefit costs to be around the same as the prior case. The next array compares total benefit costs and gross rebates as a function of market share for both position assignments without and with lump sum rebates:

    Exclusive - no lump sum rebates
    Shared - no lump sum rebates
    Exclusive -- lump sum rebates
    Shared - lump sum rebates

| Total Benefit Costs and Gross Rebates | | | | | | | |
|---|---|---|---|---|---|---|---|
| Market share - Incumbent | 90% | 85% | 80% | 75% | 72% | 70% | 65% |
| Market share - Entrant | 10% | 15% | 20% | 25% | 28% | 30% | 35% |
| | | | | | | | |
| Total Benefit Cost $M - Exclusive - no lump sums | $3,029 | $3,029 | $3,029 | $3,029 | $3,029 | $3,029 | $3,029 |
| Total Benefit Cost $M - Shared - no lump sums | $3,235 | $3,156 | $3,078 | $2,999 | $2,952 | $2,920 | $2,841 |
| | | | equal | | | | |
| Total Benefit Cost $M - Exclusive - lump sums | $3,029 | $3,029 | $3,029 | $3,029 | $3,029 | $3,029 | $3,029 |
| Total Benefit Cost $M - Shared - lump sums | $3,140 | $3,087 | $3,034 | $2,982 | $2,950 | $2,929 | $2,876 |
| | | | | | | | |
| Gross Rebates $M - Exclusive - no lump sums | $3,029 | $3,029 | $3,029 | $3,029 | $3,029 | $3,029 | $3,029 |
| Gross Rebates $M - Shared - no lump sums | $2,823 | $2,902 | $2,981 | $3,060 | $3,107 | $3,138 | $3,217 |
| | | | equal | | | | |
| Gross Rebates $M - Exclusive - lump sums | $3,029 | $3,029 | $3,029 | $3,029 | $3,029 | $3,029 | $3,029 |
| Gross Rebates $M - Shared - lump sums | $2,919 | $2,971 | $3,024 | $3,077 | $3,109 | $3,130 | $3,182 |

Given any market share, the array above shows that the same benefit costs and gross rebates received could be achieved with or without lump sum rebates. PBMs and plan sponsors do not necessarily benefit from the addition of lump sum rebates. We believe that the motive for a PBM to add lump sum rebates is weak.

On the other hand, both a PBM and a plan sponsor benefit by an entrant exceeding expected market share. Given any assignment and bases, the array above shows that an entrant's improving market share reduces total benefit costs and increases rebates due to the relative differences in unit shared bids -- (b3 - b2)

**Disclosures:**

I have not received any remuneration for this paper nor have I financial interest in any company cited in this working paper.

I have a Ph.D. in Economics from Washington University in St. Louis and a B.A. in Economics from Amherst College. Other papers on PBMs can be found on my website https://nu-retail.com

© Lawrence W. Abrams, 2023